\documentclass[longauth,letter]{aa} 

\usepackage{graphicx}
\usepackage{txfonts}
\def\degr{\hbox{$^\circ$}}

\def\fdg{\hbox{$.\!\!^\circ$}}

\def\farcs{\hbox{$.\!\!^{\prime\prime}$}}

\usepackage{color}

\begin{document} 

\title{A Tale of Three Cities}
\subtitle{OmegaCAM discovers multiple sequences in the color-magnitude diagram of the Orion Nebula Cluster}
\titlerunning{OmegaCAM discovers multiple sequences in the ONC}

\author{G. Beccari\inst{1} \and
M.G. Petr-Gotzens\inst{1} \and
H.M.J. Boffin\inst{1}\and 
M. Romaniello\inst{1,12}\and 
D. Fedele\inst{2}\and 
G. Carraro\inst{3}\and 
G. De Marchi\inst{4}\and\\
W.-J. de Wit\inst{5}\and 
J.E. Drew\inst{6}\and 
V.M. Kalari\inst{7}\and 
C.F. Manara\inst{4}\and 
E.L. Martin\inst{8}\and 
S. Mieske\inst{5}\and 
N. Panagia\inst{9}\and 
L. Testi\inst{1}\and 
J.S. Vink\inst{10}\and\\
J.R. Walsh\inst{1}\and
N.J. Wright\inst{6,11}
}
\authorrunning{G. Beccari et al.}
\institute{European Southern Observatory, Karl-Schwarzschild-Strasse 2, 85748 Garching bei M\"unchen, Germany, \email{gbeccari@eso.org}\and
INAF-Osservatorio Astrofisico di Arcetri, L.go E. Fermi 5, 50125, Firenze, Italy\and
Dipartimento di Fisica e Astronomia Galileo Galilei, Vicolo Osservatorio 3, I-35122, Padova, Italy\and
Science Support Office, Directorate of Science, European Space Research and Technology Centre (ESA/ESTEC), Keplerlaan 1, 2201 AZ Noordwijk, The Netherlands\and
European Southern Observatory, Alonso de Córdova 3107, Casilla 19001, Santiago, Chile\and
School of Physics, Astronomy and Mathematics, University of Hertfordshire, College Lane Campus, Hatfield, AL10 9AB, UK\and
Departamento de Astronomía, Universidad de Chile, Casilla 36-D, Correo Central, Santiago, Chile\and
CSIC-INTA Centro de Astrobiologia Carretera Ajalvir km 4 28550 Madrid, Spain \and
Space Telescope Science Institute, 3700 San Martin Drive, Baltimore, MD 21218, USA\and
Armagh Observatory, College Hill, Armagh BT61 9DG, United Kingdom\and
Astrophysics Group, Keele University, Keele ST5 5BG, UK\and
Excellence Cluster Universe, Garching bei M\"unchen, Germany 
}


\abstract{As part of the Accretion Discs in H$\alpha$ with OmegaCAM (ADHOC) survey, we 
imaged in $r$, $i$ and $H\alpha$ a region of $12\times8$~square degrees around the Orion Nebula Cluster. Thanks to the high-quality photometry obtained, we discovered three well-separated pre-main sequences in the color-magnitude diagram. The populations are all concentrated towards the cluster's center. Although several explanations can be invoked to explain 
these sequences we are left with two competitive, but intriguing, scenarios: a population of unresolved binaries with an exotic mass ratio distribution or three populations with different ages. Independent high-resolution spectroscopy supports the presence of discrete episodes of star formation, each separated by about a million years.
The stars from the two putative youngest populations rotate faster than the older ones, in agreement with the evolution of stellar rotation
observed in pre-main sequence stars younger than 4~Myr in several star forming regions.\, 
Whatever the final explanation, our results prompt for a revised look at the formation mode and early evolution of stars in clusters.}

\keywords{Stars: pre-main sequence --  open clusters and associations: Orion}

\maketitle
%

\section{Introduction}
\label{sec:intro}

Young stellar clusters are conspicuous components of our Galaxy. 
They are the best test beds of the stellar initial mass function
because they are assumed to be entities of a common origin, i.e.\ born from the same molecular cloud material, and at the same time. Observations have shown, however, that the stars born in one and the same cluster are not as coeval as expected. In fact, stellar age spreads up to several million years have been postulated for several clusters from fitting isochronal ages to the position of cluster member stars in the Hertzsprung-Russell diagram~\citep[HRD; e.g.][]{ca10,ci10,re11,be13,ba16}. But, if this observation is a consequence of a continuous star formation process lasting over several dynamical time scales, or rather is caused by different accretion histories within an otherwise co-eval population of pre-main sequence (PMS) stars, is widely debated~\citep{da10,je11}.

The Orion Nebula Cluster (ONC) is the nearest~\citep[414~pc;][]{men07}, populous young stellar cluster and hence the best laboratory to test the existence of stellar age spreads. \citet{Palla05,Palla07} were the first to point out an apparent large age spread of $\sim$10\,Myr based on the evidence of lithium depletion in some stellar cluster members. \citet{da10} interprets the luminosity spread of the ONC's PMS as an age spread of 0.3-0.4 dex with a mean age of 3-4\,Myr depending on the models.

In this letter we present a photometric study of a $12\degr\times8\degr$ area including the ONC. We report the detection for the first time of multiple sequences in the observed optical color-magnitude diagram of the ONC.

\section{Observations and data reduction}
\label{sec:obs}

The images used in this work were collected with the wide field optical camera OmegaCAM on the 2.6-m VLT Survey Telescope (VST) at Cerro Paranal in Chile. OmegaCAM consists of a mosaic of 32 CCDs and samples a 1 deg$^2$ Field of View (FoV) with
a pixel sampling of 0.21 arcsec pixel$^{-1}$. Orion was sampled through the $r,i$ broad band filters and the H$\alpha$ narrow band as part of the Accretion Discs in H$\alpha$ with OmegaCAM (ADHOC) survey (PI: Beccari). 
Each target region is sampled in groups of 3 overlapping fields, and in each group the fields are contiguous with a footprint close to $3\times1$ deg$^2$. For each position in sky we acquire 2 exposures of 25 sec in $r$ and $i$ and 3 images of 150 sec exposure with the H$\alpha$ filter.
The frames were collected between October and December 2015 and the median image quality is $0\farcs91\pm0\farcs16$. 

\begin{figure*}
\centering
\sidecaption
\includegraphics[width=12.9truecm]{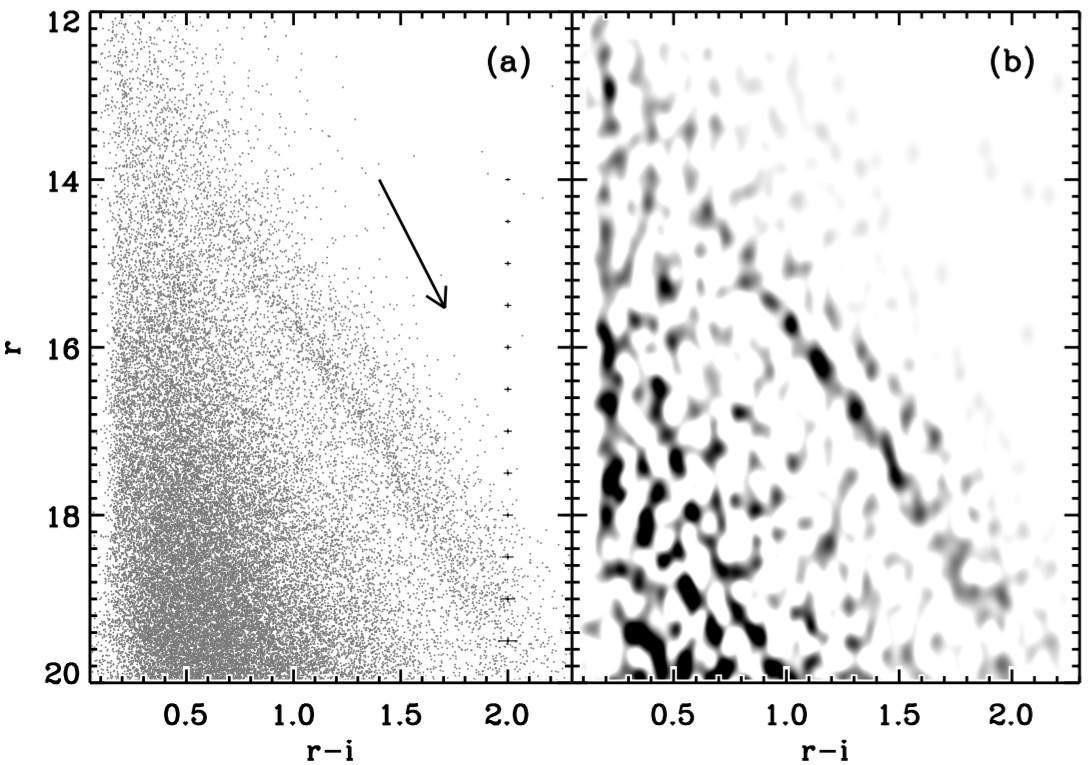}
\caption{{\it Panel a}: The (r-i),r color-magnitude diagram of a region of $1\fdg5$ radius centered on the ONC. The arrow represents a reddening vector for $A_v=1.8$~\citep[typical for Orion;][]{sch14} and extinction curve with $R_v=3.1$~\citep{ca89}. The photometric errors (magnitudes and colors) are indicated by black crosses; {\it Panel b} the same CMD shown in panel (a) after unsharp masking. \vspace{5.0cm}
}
\label{fig:cmd}
\end{figure*}
%
\begin{figure*}
\centering
\sidecaption
\includegraphics[width=11.9truecm]{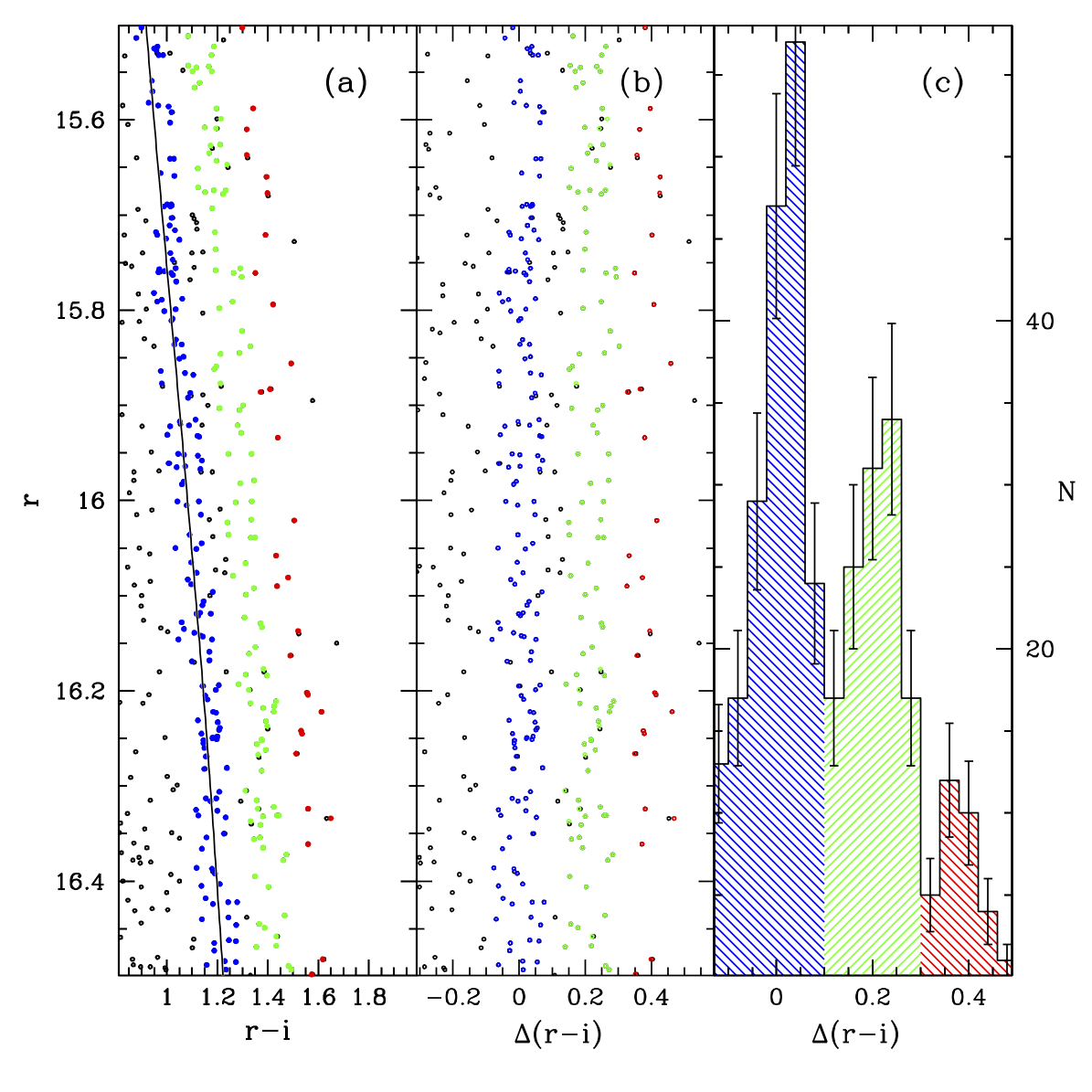}
\caption{(a) Portion of the CMD zoomed on the PMS. The black line shows the mean ridge line of the blue population; (b) Rectification of the CMD shown in panel a; (c) Histogram of the distance in (r-i) color of the PMS stars
from the mean ridge line of the bluest population. 
\vspace{9.7cm}
}
\label{fig:hist}
\end{figure*}
%
The pattern to these observations is similar to that of the VPHAS+ survey, and the data have passed through the same pipeline~\citep[see][for more details]{dr14}. The entire data-set was fully processed, from the bias, flat-field and linearity correction to the stellar photometry,
at the Cambridge Astronomical Survey Unit (CASU).
The magnitude for each star is extracted using aperture photometry adopting
an algorithm based on IMCORE\footnote{Software publicly available from http://casu.ast.cam.ac.uk.}~\citep[][]{ir85} and the
nightly photometric calibrations are also performed. 
We downloaded the astrometrically and photometrically calibrated single band 
catalogs from the VST archive at CASU\footnote{http://casu.ast.cam.ac.uk/vstsp/}. 
Stars lying in the overlap region between adjacent fields were used
to adjust residual photometric offsets. The photometric calibration of the final band-merged catalog covering the entire area was checked against a catalog of stars from the AAVSO Photometric All Sky Survey, used as $secondary~ standard$ catalog.  
The final catalog allows us to homogeneously sample the stellar populations in a region of $12\degr\times8\degr$ size in Orion down to $r\sim20$.

\section{Three distinct pre-main sequences in the ONC}

We show in Fig.~\ref{fig:cmd}a the ($r-i$) vs. $r$ color-magnitude diagram (CMD) of the stars located inside a radius of $1\fdg5$
from the center of the ONC. 
The population of PMS objects is well detected 
and occupies the reddest side of the CMD above the population of back/fore-ground stars in the magnitude 
range $14<r<20$ and $(r-i)>1$, which roughly corresponds to PMS stars of masses between 0.2 and 1 M$_{\odot}$~\citep[see also][]{da09,da16}. 
A remarkable feature is well visible on the CMD shown in Fig.~\ref{fig:cmd}a, i.e.
{\it the presence of at least two distinct and near-parallel sequences of PMS stars}. 
Following~\citet{de16}, in Fig.~\ref{fig:cmd}b we have applied to the CMD the ``unsharp-masking'' technique of~\citet{sp31} in order to enhance the high-frequency features in the diagram. The result shown in Fig.~\ref{fig:cmd}b further supports the detection of at least two distinct PMSs in the CMD diagram. 
We stress here that by inspecting the entire surveyed area we found that this feature is clearly detectable only in the ONC region.

We used a simplified version of the method described in~\citet{mi09} to further investigate the existence
of two or more distinct sequences in the CMD. In Fig.~\ref{fig:hist}a we show a portion of the CMD
zoomed on the population of the PMS. In order to increase the contrast of the ONC population against the back/fore-ground stars, we show only stars inside a radius of $0\fdg5$ from the center of the cluster. The black line shows the mean ridge line of the blue PMS. We then calculate
the distance in $r-i$ color of each star in a magnitude range $15.5<r<16.5$ from the
mean ridge line arbitrary chosen as reference line.
We adopted this magnitude range to provide an adequate statistical
balance between the number of stars in the PMS population and 
the contamination from field stars. The distance of each star 
as a function of the $r$ magnitude is shown on panel (b) of the same figure.
Panel (c), the histogram of the distances in ($r-i$) color, clearly shows the presence
of even three distinct populations of PMS stars well separated in color on the CMD. Indeed the Hartigans' dip test confirms that the distribution of the color distances is incompatible with a uni-modal's one. The number fraction of stars belonging to the two populations with the reddest colors (green and red in the figure)  compared to the reference one (blue) is 0.5 and 0.15, respectively. These fractions hold (within the statistical uncertainties) even after accounting for contamination, which we estimated to be between 15-40\% based on the CMD of a control field located a few degrees west of the ONC, and spectroscopic membership information from APOGEE spectra (see Sec.~\ref{sec:d16}). 

\begin{figure*}[htbp]
\begin{center}
\includegraphics[width=15cm]{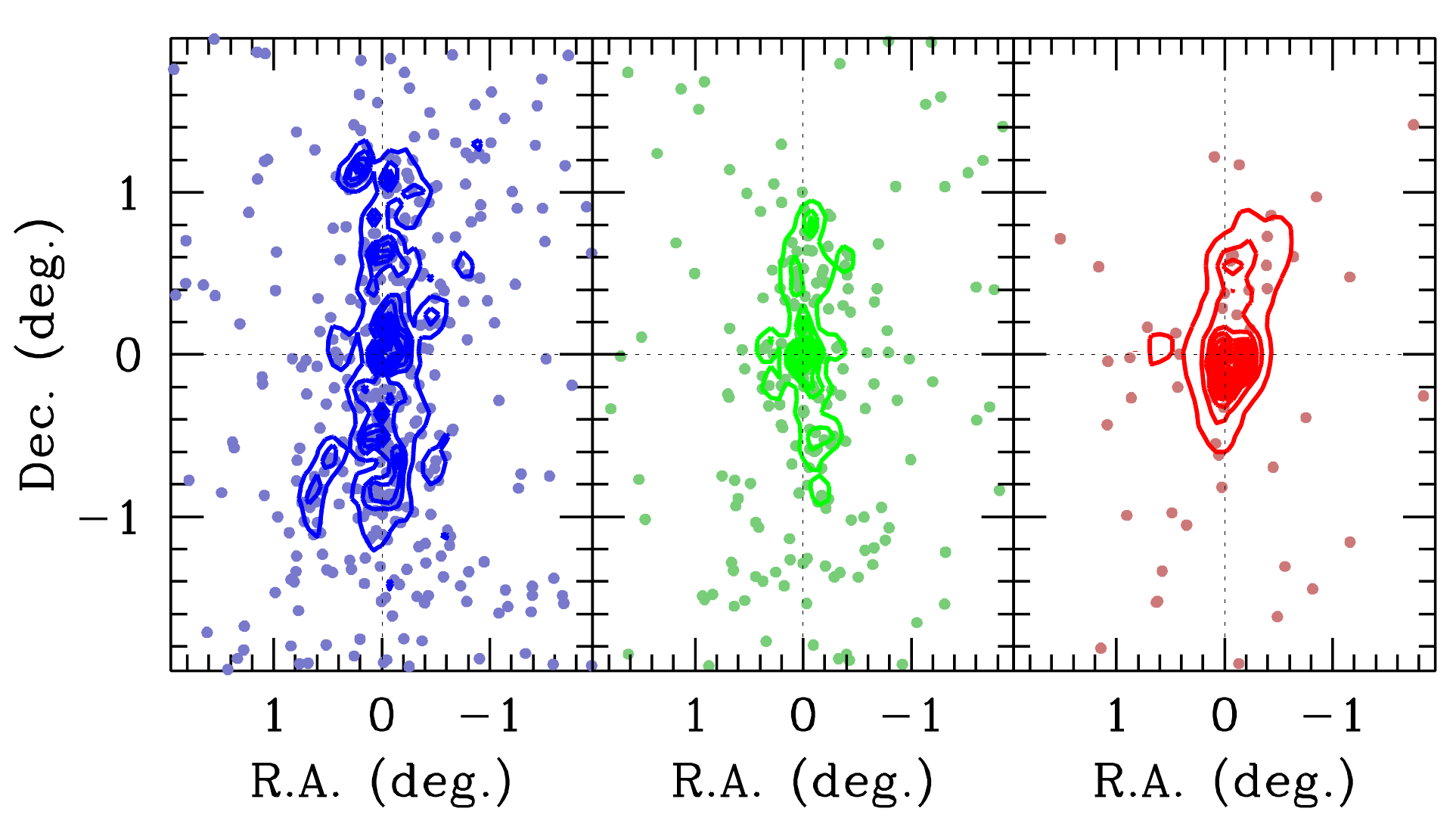}
\caption{\label{Fig:densmap} Surface density of the three populations, old (in blue), young (in green), and very young (in red) together with the position of the stars. The plots have been centered on the ONC nominal center. All contours are normalized to the maximum value of the population itself. The location of the stars belonging to each population (solid dots) are also shown.
}
\end{center}
\end{figure*}

Next, we investigate the spatial 
distribution of stars belonging to the different sequences. In Fig.~\ref{Fig:densmap} we show the surface densities of the three populations. These density plots are calculated using all the stars belonging to the three PMS populations selected in the color and magnitude ranges shown in the CMD of Fig.~\ref{fig:hist}. We calculate the densities using the {\tt GATHER} method from~\cite{1999MNRAS.302..305G}.
In the maps of Fig.~\ref{Fig:densmap} we show the position of the stars with respect to the ONC nominal center (solid circles) together with the density contours, which were scaled to the maximum value of each population. It emerges that the density distributions of the three populations all peak around a common center. The blue population seems to be slightly more sparsely distribute with respect to the spatial distribution of the stars belonging to the green and red ones.

We used the Minimal Spanning Tree~\citep[MST; see][]{al09} to assess if any difference in the spatial distribution of the populations is present and at which level of significance. The MST is the unique set of straight lines ("edges")
connecting a given sample of points ("vertices"; in this case the star
coordinates) without closed loops, such that the sum of the edge
lengths is the minimum possible.  Hence, the length of the MST is a
measure of the compactness of a given sample of vertices~\citep{cw04}. The MST is a powerful algorithm to study population distributions since it can be used without assuming that the studied populations are distributed around the same center of gravity, which is mandatory when using the Kolmogorow-Smirnov (KS) test. The weakness of this method is that it must be assumed that the photometric completeness of the compared populations is the same. We have selected the populations in a common range of magnitudes that are affected by the same level of completeness. The fact that the three populations are affected by the same contamination from field stars, makes it not possible to use the MST when the number of genuine members is low, which is the case for the red population. For this reason we limit the use of the MST to the blue and green populations.

We first estimate the MST of all the stars in the green population. Then we randomly extract 1,000 sets of stars belonging to the blue population equal in number to the number of stars in the green one. We compute the MST of the blue populations as the mean and the standard deviation of the distribution of 1,000 MSTs. Hence, we calculate\footnote{We remind here that $\Lambda MST=1$ means that the compared populations are equally distributed in space while a value greater than one indicates that the population at the numerator is more spatially extended then the one at the denominator.} 
$$ \Lambda MST_{blue}= \frac{MST_{blue}}{MST_{green}}.$$
We find that $\Lambda MST_{blue}=1.16\pm0.03$, which indicates that the blue sequence is slightly more sparsely distributed with respect to the green one with a 5$\sigma$ significance. 

\section{The unresolved binaries/multiples scenario}

The appearance of multiple parallel sequences in the CMD as shown in Fig.~\ref{fig:cmd} could be caused by
(1) a single coeval population consisting of single stars and unresolved
binaries and higher order multiples; or populations (2) at different
distances; (3) with different extinction A$_V$; or (4)  with different ages. In this and the following sections we will discuss these four hypotheses.

Any unresolved binary system would appear in a CMD as a single star with a flux equal to the sum of the
fluxes of the two components. This effect produces a systematic over-luminosity of these objects and a shift in color which depends on the magnitudes (and hence mass) of the two components in each passband. When the mass ratio of two stars in the binary system is $q=1$ (equal mass binary) 
the unresolved binary system will appear 0.752 mag
brighter than the individual component's brightness. The peak of the distribution of the green population is $\sim0.75$mag brighter than the blue one (Fig.~\ref{fig:cmd}a) and this distribution may therefore well reflect the presence of a population of unresolved ONC binaries. 
Under the assumption that the green and red sequences represent the binaries and higher-order multiples of the ONC members, we derive a multiplicity fraction of ~39\%. This number would support the unresolved binary hypothesis given that this fraction overall agrees with that seen for low-mass stars in other young clusters~\citep{luh05,rei14} and with the multiplicity fraction among field M-dwarfs~\citep{dk13}.

\begin{figure*}[htbp]
\centering
\sidecaption
\includegraphics[width=12.9truecm]{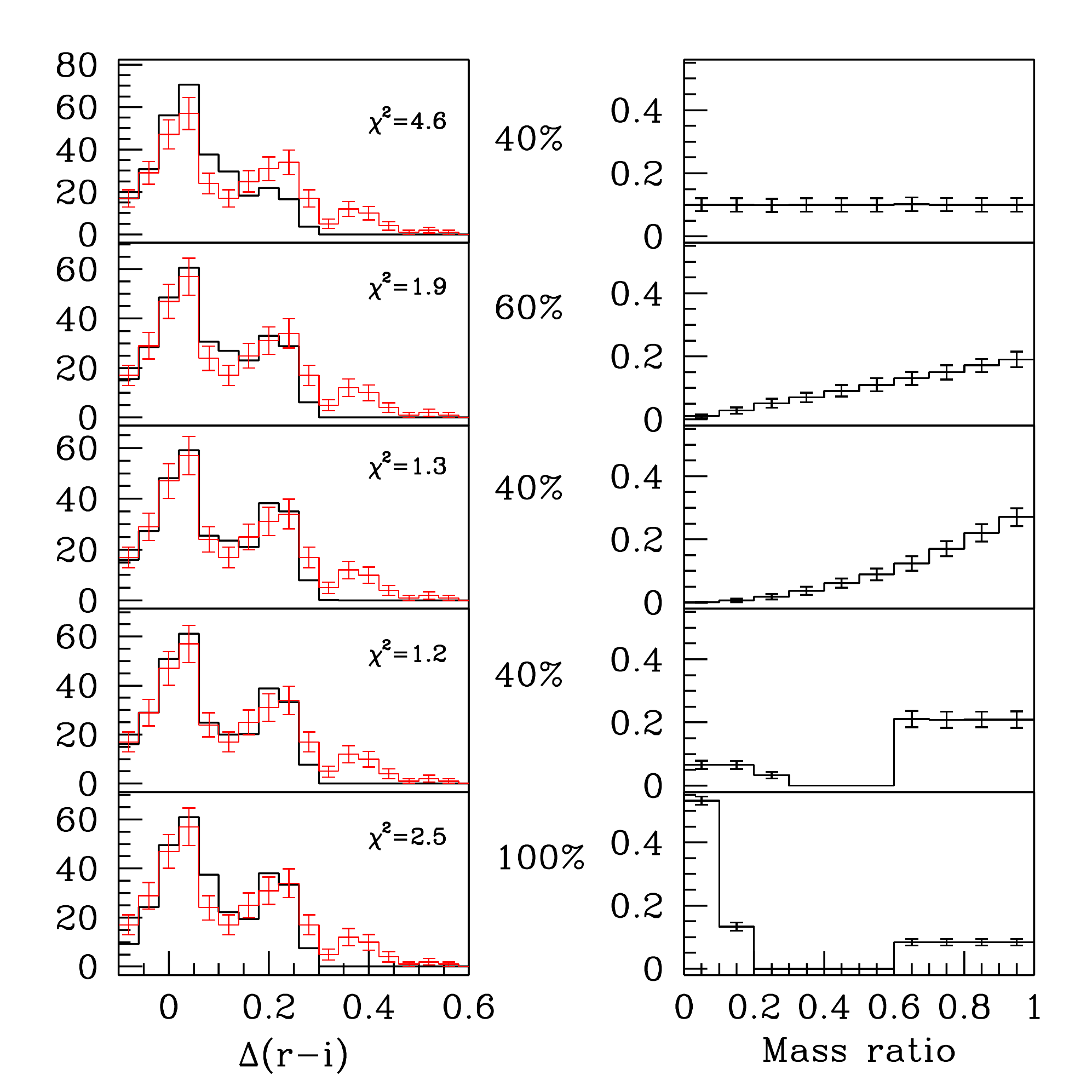}
\caption{\label{fig:fq} The left panels show the histograms of the distance in the (r-i) color of the PMS stars from the mean ridge line of the bluest population, as obtained with our Monte Carlo simulations (black, heavy lines) and as observed (red). The right panels show the corresponding assumed distribution of the mass ratios of all the binaries in our simulated sample. The $\chi^2$ of each solution is also indicated in the left panels. Both top panels correspond to the canonical cases of 40\% multiplicity and a uniform mass ratio distribution.
\vspace{6.99cm}}
\end{figure*}

Beside the total fraction, the companion mass ratio distribution plays an important role in shaping the CMD appearance. Determinations of the mass ratio distribution for ONC binaries are rare. Those studies that have
derived mass ratios for close visual systems and spectroscopic systems find no indication for an equal mass
preference~\citep[][]{da12,co13,ko16}, i.e.\ for those 
systems that lead to the largest displacement in luminosity in the CMD. 
At most, the observed mass ratio distribution slightly increases from low~$q$ to higher $q$~\citep[][]{co13}, although this result is most likely affected by incompleteness and selection effects at low $q$.
\citet{w15} performed a companion survey of 245 late-K to mid-M (K7-M6) dwarfs within 15 pc and found that the mass ratio distribution across the $q=0.2-1.0$ range is flat. This seems to be a general result. Numerical simulations of primordial binaries~\citep[e.g.][]{ba09} produce an $f(q)$ that is rather flat too. Moreover $f(q)$ seems to be rather insensitive to dynamical disruptions and interaction processes within the cluster~\citep[][]{pr13}.

The marked CMD morphology and in particular the presence of the two gaps
in the distribution of colors shown in Fig.\ref{fig:hist}c, allows us to investigate which combination
of total binary frequency and companion mass ratio distribution could explain the observed CMD, and which
can be excluded.
For this, we performed Monte Carlo simulations and test a range of total binary fractions and mass ratio
distributions $f(q)$\footnote{Note that when doing this, we only use the blue and green populations, and ignore for now the red population.}. We draw randomly a star from the blue sequence (with its $r$ and $i$ magnitudes) and, using the mass-luminosity relation from~\citet{bre12}, we determine its mass. We then draw a mass ratio from a given $f(q)$ distribution, and add a companion with a mass $q~m_1$. We then compute the color and magnitude of the resulting binary and can then see where the so obtained binary falls in the CMD. Finally, we compare the obtained histograms of colors with that observed in Fig.\ref{fig:hist}c. A range from 40\% to 100\% in binary fraction and different $f(q)$ of the form uniformly flat, linearly increasing, quadratically 
increasing, or step-like were explored. In Fig.~\ref{fig:fq} we show a few representative results.

It is clear that the canonical case of 40\% binary fraction with a uniform mass-ratio distribution, as shown in the upper
panels of Fig.~\ref{fig:fq}, does not provide a satisfactory agreement with the observations. Increasing the 
fraction of binaries even worsens the comparison with observations, as does a lower total binary fraction.
We conclude that a uniformly flat mass ratio distribution is not able to reproduce the observations, in particular
it is not capable in reproducing the obvious, significant gaps in the CMD. 
When assuming other mass ratio distributions we can achieve reasonable fits (e.g.\ second to fourth row panels in Fig.~\ref{fig:fq}).
Here we considered as reasonable fit any solution with a reduced $\chi^2 < 2$ which indicates that it can be 
trusted with 99.5\% confidence. The solutions come with some caveats, though. A linearly 
increasing $f(q)$ (second row panels of Fig.~\ref{fig:fq}) appears only possible in combination with a total 
binary fraction of around 60\%. However, such high overall binary frequencies
among ONC low-mass stars are not observed. For visual binaries with separations between 
a few tens and a few hundreds of AU (which would indeed appear unresolved in our OmegaCAM observations) various
studies have consistently found that the ONC binary fraction is even slightly {\it lower} than in the field~\citep[][]{pe98,ko06,re07}. Also very close binaries with separations $<$10\,AU, that have been traced by a multi-epoch spectroscopic study~\citep[][]{to13,ko16} do not show any excess in companions, but are again consistent with field star fractions. Hence, the overall binary fraction cannot noticeably exceed 40\% unless these excess binaries are all in the narrow separation range of $\sim10-40$\,AU which 
seems very unlikely.
On the other hand, if we enforce mass ratio distributions that are strongly skewed towards high mass ratios, such as a quadratically increasing or step-like function where the majority of systems would have $q>0.6$, then a simulated population with a total binary fraction
as low as 35-40\% can reproduce the observations (third and fourth row panels of Fig.~\ref{fig:fq}). Actually, also higher total binary fractions, up to 65\% would be feasible according to our simulation. In this case, however, the blue population would contain a large number of unresolved low $q$ systems. Such specific mass ratio distributions are at odds with observations.
We are thus led to the conclusion that in order for the binary hypothesis to be valid, one would need to postulate a very distinct mass-ratio distribution. 
 
In the same vein, the reddest sequence we find in the CMD would be even more difficult to explain, as it would require triple systems, where, the mass ratio distributions would have also to be quite tailored, i.e. the mass ratios of both secondaries and tertiaries would need to be strongly peaked to $q > 0.8$ and be exactly similar.

Although formally we cannot exclude the presence of a population of binaries with an unusual $f(q)$, the fact that no young binary population in any star forming region has shown indications of such an usual $f(q)$, casts some
doubts on binaries as the origin for the observed multiple sequences. On the other hand,
if confirmed this fact would certainly challenge many of the studies published so far on the stellar population in the ONC whenever such stars were considered as single objects~\citep[e.g.][]{hi13,da16}.

\section{Differential A$_{V}$ and distances scenario}

Any effect of differential extinction can be rejected by the fact that the reddening vector as shown on the CMD of Fig.~\ref{fig:cmd}a runs parallel to the PMSs.

\begin{figure}[htbp]
\includegraphics[width=8cm]{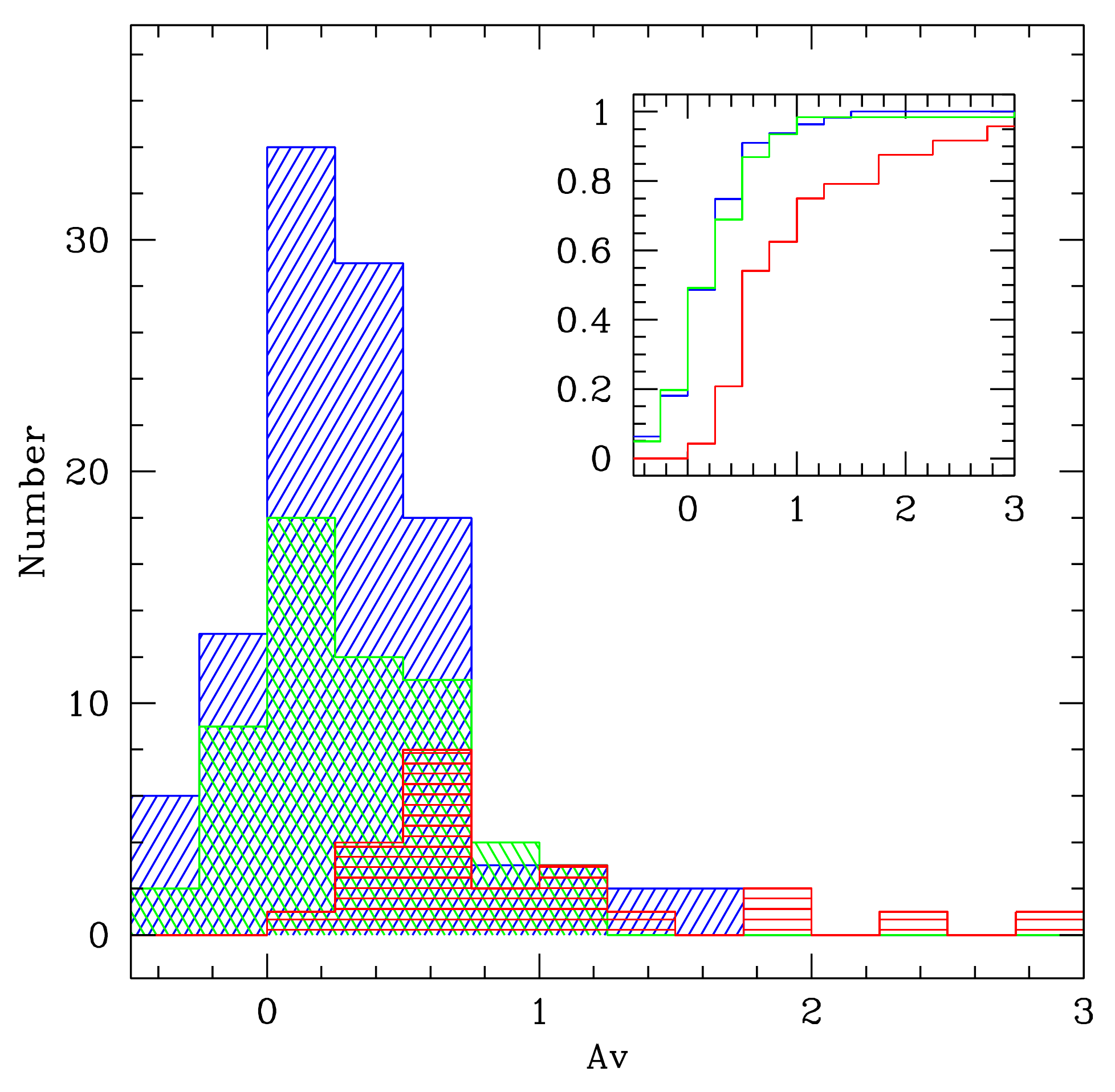}
\caption{\label{Fig:Av} Distribution of the visual extinction, $A_v$, for the stars from the three populations, old (in blue), young (in green), and very young (in red). The inset shows the normalized cumulative fraction. The Kolmogorov-Smirnov test indicates that the $A_v$ distribution of the old and intermediate-young population are extracted from the same parental distribution with more than 5$\sigma$ significance. The very young population shows hints of a slightly higher $A_v$ distribution but is affected by low number statistics.}
\end{figure}

Moreover, as shown in Fig.~\ref{Fig:Av} we have verified that the stars from the three sequences have approximately the same distribution of visual extinction A$_V$. To this aim we adopted the A$_V$ provided by the spectroscopic analysis of~\citet[][]{da16}. 

An alternative explanation for the presence of the green and red sequences could be a PMS population located more than 100 and 200 pc in the foreground of the blue sequence, respectively. \citet{al12} suggested that the population of PMS around $\iota$~Ori belongs to the association NGC~1980 and represents a population of 4-5 Myr old stars located at $\sim30$~pc in the foreground of Orion A~\citep{bou14}. In their detailed spectroscopic analysis~\citet{da16} rule out this hypothesis finding that the candidate foreground population is kinematically indistinguishable from the Orion A's one. They conclude that the old population studied by~\citet{bou14} witnesses the earliest (i.e. oldest) episode of star formation in the ONC. The three distinct PMS populations found in our CMD and their spatial distribution are not compatible with the presence of a old foreground population of PMS stars. ~\citet[][]{fa17} performed and extensive spectroscopic analysis of 691 ``foreground'' stars in the Orion A region and confirm that NGC~1980 is not a foreground population.
Considering that the candidate foreground population studied by~\citet{bou14} is located more than 1~deg south with respect to the peaks of the density maps of the three populations shown in Fig.~\ref{Fig:densmap} we think that it is unlikely to be related.

\section{Three discrete episodes of star formation}
\label{sec:d16}

In the previous sections we show that differential extinction or distance offsets are very unlikely to explain the discovered features in the CMD. The presence of unresolved binary and tertiary populations 
reproduce the CMD morphology only if the underlying mass ratio distribution is rather unusual (Fig.~\ref{fig:fq}).

We here explore the possibility that the age is the origin of the discreetness of the color distribution of the PMS in the ONC. We verified that the distance in magnitude between a 1~Myr and a 3~Myr PMS isochrone from~\citet[][]{bre12} in the $r, r-i$ CMD  is indeed $\sim0.75$, i.e. equal to the the shift in luminosity due to unresolved binaries. This unfortunate fact is at the root of why it is hard to distinguish between the binary hypothesis and the multiple population scenarios\footnote{And we have checked that using different combinations of filters wouldn't help in this respect.}.

In order to assign ages to the stars belonging to the three distinct PMS populations, we use the measurements presented by~\citet[][D16 hereafter]{da16}. D16 performed a spectroscopic study of the young stellar population of the Orion A molecular cloud with the APOGEE spectrograph. In their work they measured accurate stellar parameters (${T}_{{\rm{eff}}}$, $\mathrm{log}g$, $v\mathrm{sin}i$) and extinctions and conclude that star formation in the ONC proceeded over an extended period of $\sim3$ Myr age. 
We have cross-correlated our stars with the catalog of stellar parameters (including age, temperature, extinction) published in table 4 of D16. We used these stellar parameters to study the
properties of the three candidate populations of PMS stars as selected in Fig.~\ref{fig:hist}. In order to have a sample which is as clean as possible from any contamination we removed stars from their study that did not clearly obey a PMS mass-luminosity relation and that, based on the effective temperature, mass and luminosity, cannot populate the PMS region in the HRD. We were left with 111, 63, and 24 stars for the three sequences (blue, green, and red), respectively.

\begin{figure}[htbp]
\includegraphics[width=8cm]{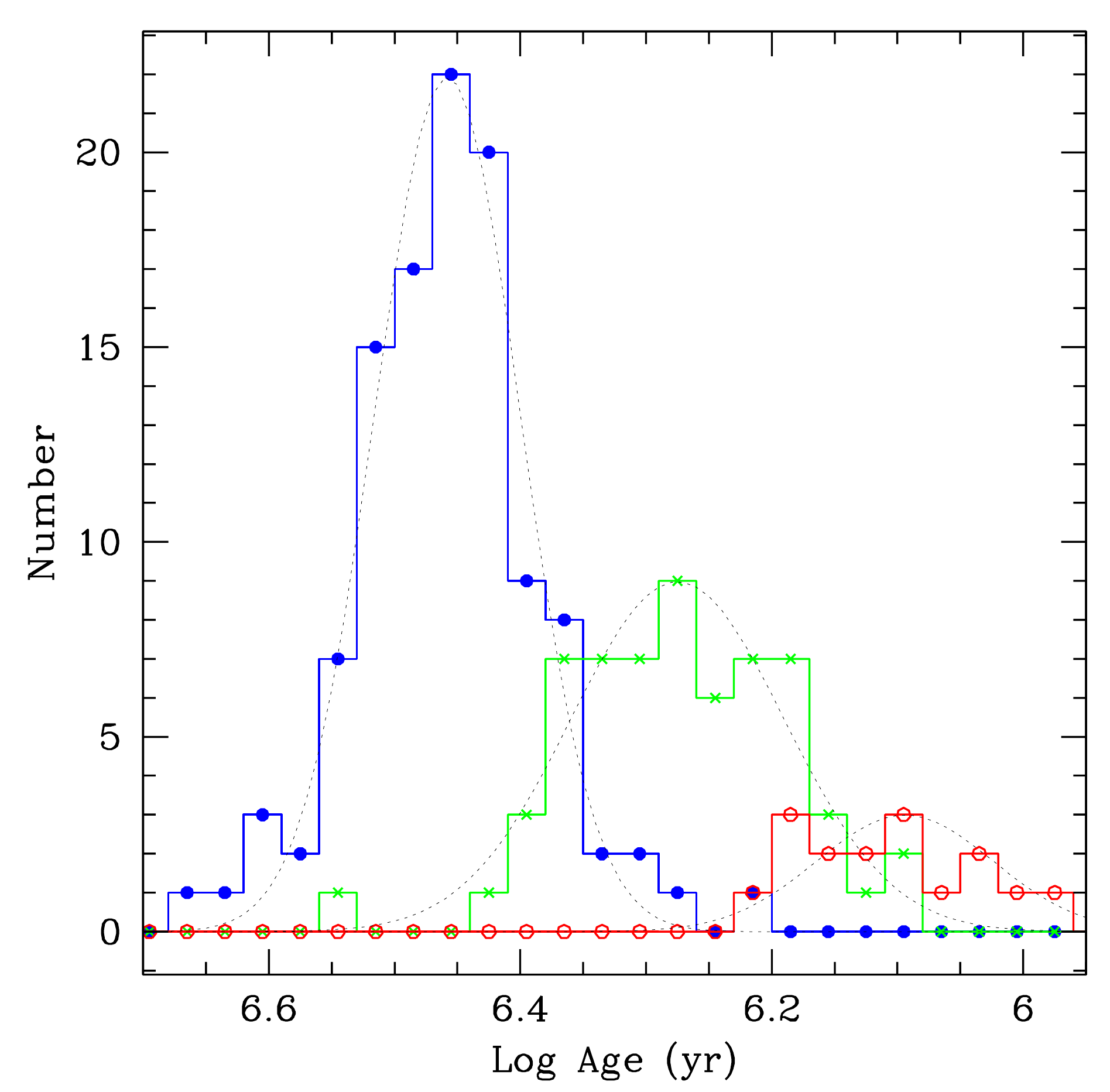}
\caption{\label{Fig:age2} Distribution of the logarithm of the spectroscopically determined ages for stars from the three populations, old (in blue), young (in green), and very young (in red) with the best Gaussian fits indicated.
}
\end{figure}

\begin{table}[htbp]
    \caption{  \label{tab:pops} Properties of the three populations.}
   \begin{tabular}{@{} lccc @{}} 
        \hline\hline
 & {\small old}& {\small young}& {\small very young}\\
\hline
{\small Mean Log age (yr) }& {\small 6.46$\pm$0.06} & {\small 6.27$\pm$0.09} & {\small 6.09$\pm$0.07}\\
{\small Mean age (Myr) }& {\small 2.87} & {\small 1.88} & {\small 1.24}\\
{\small 1-$\sigma$ age interval (Myr) }& {\small 2.51--3.28} & {\small 1.55--2.29 }& {\small 1.08--1.53}\\
{\small 5--95\% interval (Myr) }&   {\small 2.30--3.58 }& {\small 1.37--2.60} & {\small 1.04--1.63}\\
{\small Rotational velocity (km/s) }& {\small 14$^{+6}_{-4} $} & {\small 25$^{+25}_{-12}$} &  {\small 35$^{+36}_{-16}$} \\
\hline
\end{tabular}
\end{table}

In Fig.~\ref{Fig:age2} we show the distributions of the ages derived by 
D16 of the stars in the three samples. These age histograms indicate that the three sequences that we discovered in the photometric study have distinct ages, with the bluer population being the oldest. We have fitted the distributions with Gaussians, using a $\chi^2$-minimization technique, to derive the mean and standard deviation ($\sigma$). As these are distributions of the logarithm of the ages, we then estimated the corresponding distributions of the ages and computed the respective 1-$\sigma$ and 5--95\% intervals. Our results are shown in Table~\ref{tab:pops}. 

{\bf This finding is consistent with the hypothesis that the formation of the population of PMS stars in the ONC that was thought to be the outcome of a single episode extended over 3 Myr is instead best
described by 3 discrete and sequential episodes of star formation over the same time span.} 

\begin{figure}[htbp]
\includegraphics[width=8.3cm]{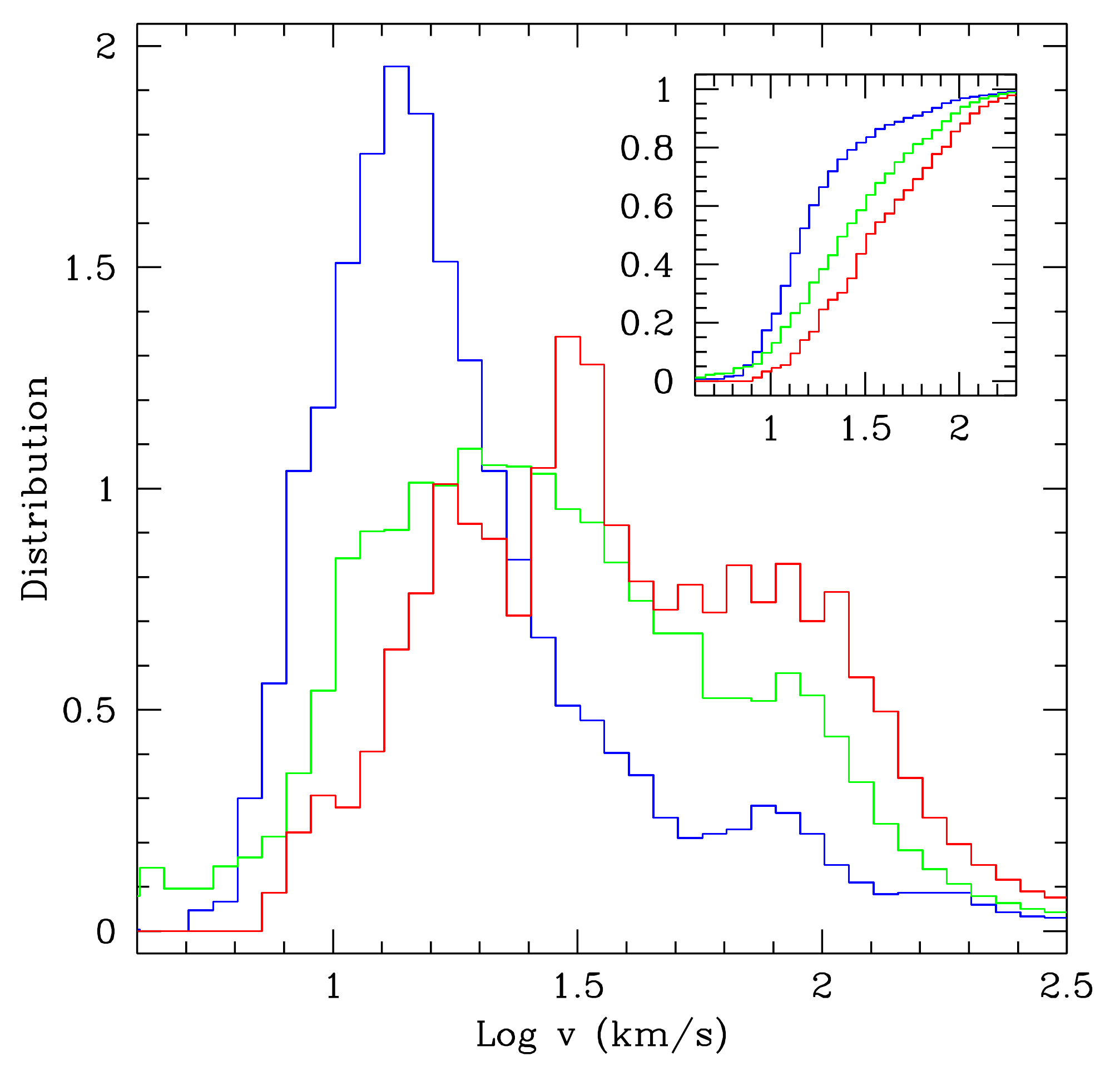}
\caption{\label{Fig:vdist} Distribution of the logarithm of the rotational velocity of the stars from the three populations, old (in blue), young (in green), and very young (in red). The inset shows the normalized cumulative fraction. The Kolmogorov-Smirnov test indicates that the distribution of rotational velocity of the old and younger populations are not extracted from the same parental distribution with more than 5$\sigma$ significance. The same applies when the non-deconvolved rotation
velocities are used.}
\end{figure}

D16 provide also stellar rotational velocities. However, spectroscopy only provides $v \sin i$, where $i$ is the unknown inclination of the rotation axis on the plane of the sky. As we may assume that $i$ is isotropically distributed in the sky, it is possible to deconvolve, using a Richardson-Lucy method \citep[see][]{1993A&A...271..125B}, the $v~\sin i$ distributions to obtain the distributions of $v$. These are shown in Fig.~\ref{Fig:vdist} and Table~\ref{tab:pops}, where {\bf it is clear that not only do the three populations have different ages, but they have also different rotational velocities}: the younger the population, the faster its members rotate. 

Obviously it can be argued that the age and $v \sin i$ estimations provided in D16 are incorrect were the redder stars entirely populated by unresolved binaries belonging to the main population. 
We investigated what it would need to obtain a broadening of the spectral line profile corresponding to v=25 km/s (the peak of the young/green population) starting from a binary system containing 2 stars having v=14 km/s (the peak of the old/blue population). It appears that this can only be explained if the binaries have orbital periods between $\sim2$ and $\sim12$ months, i.e. a very narrow range of periods. Indeed,  if the orbital period is greater than $\sim 1$ year then the maximum difference in velocities between the 2 stars is too small and cannot widen the line and make it appear as a fast rotating star. If, on the other hand, the orbital period is less than 2 months the difference in velocity is such that a double peaked line profile is detected and should have been well visible in the APOGEE spectra used by D16. 

The proponents of the binary hypothesis could also argue that the difference in rotational velocities would be due to tidal effects, as they would spin-up the stars. However, this can only work if the stars were very close to each other, i.e. with an orbital period of a few days or less \citep[e.g.][]{1992btsf.work..155M}. 
Thus, whatever explanation one chooses, one needs to have a very specific orbital period distribution, unlike what is observed for such stars. 
Coupling this fact with the narrow distribution of f(q), makes the overall multiplicity scenario very unlikely.

On the other hand, in their seminal work,~\citet[][]{reb04} studied the evolution of  periods and projected rotational velocities, $v~\sin i$,   of young 
PMS stars in the range K5$-$M2 in several star-forming regions, including the ONC.
They find a decrease in mean  $v~\sin i$ as a function of age. They conclude that a
significant fraction of all PMS stars must evolve at nearly constant angular velocity during the first 3--5 Myr. 
Several studies have been performed to look for a suitable explanation of this behavior which seems
to be related to a process of disk locking that might regulate the PMS star angular momentum during the early evolution.
In short, stars with accretion discs are found to be rotating at much slower rates, suggesting that significant angular momentum
removal mechanisms must operate during the first few Myr of formation~\citep[e.g.][]{da14, ve17}.  However, this scenario
has not yet found a general consensus~\citep[e.g.][]{he05,ci06}. 

In the context of this paper, we notice that our result on stellar rotation is not unexpected and would support what was already shown
by~\citet[][]{reb04}. We will investigate the interesting scenario of a relation between stellar rotation and ongoing
accretion from a circumstellar disk in a dedicated paper (Beccari et al. in preparation).

\section{Conclusion}

We have presented a wide-field optical survey of the stellar population in a region of $12\degr\times8\degr$ in Orion. Our CMD shows the presence of at least two parallel
sequences among the PMS stars. The distribution of the $r-i$ colors in the range
of magnitude $15.5<r<16.5$ reveals the presence of even three populations.
We investigate the origin of these sequences that were newer observed before. 

We use detailed information (including age, $A_v$, rotational velocities) published in the spectroscopic work
 of~D16 and look for comparative properties of the three populations. We can 
 safely exclude that differential extinction or different projected distances
 could be responsible for the feature revealed by our CMD.
We are hence left with two competitive, but as intriguing, explanations: 
a population of unresolved binaries or three populations with different ages.

We find that a flat mass ratio distribution is not able to reproduce the observed
CMD, regardless the binary fraction. On the other hand, we $can$ reproduce the observed color distribution shown in Fig~\ref{fig:hist} by assuming a 35\% to 65\% of binaries and a $f(q)$ strongly skewed towards high mass ratio where the majority of the binaries in the ONC have mass ratios $q>0.6$. This result, if confirmed, would provide the first and most solid constraint on the nature of compact binaries in a population of PMS objects. A dedicated spectroscopic monitoring campaign is urgently required to constrain the multiplicity fraction among these populations. This will certainly allow to unequivocally disentangle between the two scenarios. The fact that such a population of binaries populates the CMD of the ONC would inevitably challenge many of the previously published studies of stellar populations in the ONC.

The comparison with the spectroscopic measurements published by~D16 provide convincing observational evidences in support
to the hypothesis that the ONC contains three populations of PMS stars, with different ages and rotational velocities. In particular we find that the younger the population the larger is the mean rotation velocity. 

This result seems not to be unexpected. ~\citet{reb04} already reported a decrease of $vsin~i$ in the first 5~Myr
among PMS stars in the same spectral range studied in this paper and in several star forming regions, including Orion.
It has been speculated that the evolution of the angular momentum in solar-type PMS objects might be regulated
by actively accreting circum-stellar disks trough a mechanism of disk-locking. This process would 
impact the rotational properties of young stars and influence their rotational evolution~\citep[][]{da14}. 
In support of this hypothesis~\citet{re06}, using Spitzer mid-IR data for about 900 stars 
in Orion in the mass range $0.1{-}3 ~M_\odot$ find that slowly-rotating stars are indeed more likely to posses disks than 
rapidly-rotating stars. There is no agreement in the literature on this matter and our data-set will 
be used in a dedicated paper to further investigate any connection between stellar rotation and ongoing accretion.

While the unresolved binary hypothesis cannot be ruled out, the evidence described so far seem to point toward 
the first detection of distinct generations of PMS stars in the ONC.
Interestingly, using \textit{Hubble Space Telescope} (HST) observations of the ONC,~\citet{re11} found that the youngest stars in the cluster are more clustered towards the center, while the oldest ones are distributed almost homogeneously in space. Their figure 12 strongly resemble the populations' distribution shown in our Fig.~\ref{Fig:densmap}. 

~Such scenario has further interesting implication. 
In the context of investigating the origin of 'blue hook' stars in the globular cluster $\omega$~Cen, ~\citet{ta15} 
predicted that these evolved stars would originate from the evolution of a rapidly rotating second-generation PMS population whose accretion discs suffered an early disruption in the dense environment of the cluster's central regions. The result shown in this work might be pointing towards the first observational evidence that such a mechanism takes place in the early stage of a cluster's formation.

Since the centers of the spatial distributions of the 3 populations are not significantly different, we speculate that star formation has been progressing along the line of sight and that the youngest population formed, on average, further away from us. This view seems consistent with the fact that the A$_{\rm V}$ distributions for the stars in the different populations are not significantly different, but the
youngest population showing a lack of very low A$_{\rm V}$ sources. The clear decrease of number of stars belonging to the younger populations, with respect to oldest one, indicates that the overall cluster formation process is coming to an end, and that the major activity took place in the beginning when the cluster started to form. This is opposed to the model of accelerated star formation proposed by~\cite{ps00}.
In fact, the 1$\sigma$ age intervals from Table~\ref{tab:pops} indicate that the typical age spreads of the three populations are 
$\sim$0.5-0.8\,Myrs, which is in excellent agreement with the characteristic dynamical time-scale of $7\times10^5$ yr for the ONC
derived by~\citet{tan06}. Clearly, this prompts for a revised look at the formation time-scales of stars in clusters. 
%

\begin{acknowledgements}
Based on data collected through ESO programme 096.C-0730(A). This research was made possible through the use of the AAVSO Photometric All-Sky Survey (APASS), funded by the Robert Martin Ayers Sciences Fund. C.F.M. acknowledges the ESA Research Fellowship. DF acknowledges support from the Italian Ministry of Education, Universities and Research project SIR (RBSI14ZRHR). EM was supported by the Spanish Ministry of Economy and Competitiveness (MINECO) under the grant AYA2015-69350-C3-1. AllWISE makes use of data from WISE, which is a joint project of the University of California, Los Angeles, and the Jet Propulsion Laboratory/California Institute of Technology, and NEOWISE, which is a project of the Jet Propulsion Laboratory/California Institute of Technology. WISE and NEOWISE are funded by the National Aeronautics and Space Administration.

\end{acknowledgements}


\end{document}